\documentclass[10pt, final, conference, letterpaper]{IEEEtran}
\usepackage[left=.6in,right=.615in,top=.719in,bottom=.99in]{geometry}

\usepackage{amsmath,amssymb,dsfont,stfloats,color,url,bbm}
\usepackage[pdftex]{graphicx}
\usepackage[caption=false,font=footnotesize]{subfig}
\usepackage{mathabx}

\DeclareGraphicsExtensions{.eps,.pdf,.png,.jpg,.gif,.jpeg,.pstex}

\newtheorem{example}{Example}

\usepackage[noadjust]{cite}
\usepackage{url}

\newcommand*{\Resize}[2]{\resizebox{#1}{!}{$#2$}}%

\newfont{\bbb}{msbm10 scaled 700}

\newfont{\bb}{msbm10 scaled 1100}
\newcommand{\CC}{\mbox{\bb C}}

\newcommand{\EE}{\mbox{\bb E}}

\newcommand{\HH}{\mbox{\bb H}}

\newcommand{\yy}{\mathbbm{y}}

\newcommand{\zz}{\mathbbm{z}}
\newcommand{\sss}{\mathbbm{s}}

\newcommand{\hh}{\mathbbm{h}}

\newcommand{\vvv}{\mathbbm{v}}

\newcommand{\ev}{{\bf e}}

\newcommand{\hv}{{\bf h}}

\newcommand{\nv}{{\bf n}}

\newcommand{\vv}{{\bf v}}

\newcommand{\yv}{{\bf y}}
\newcommand{\zv}{{\bf z}}
\newcommand{\zerov}{{\bf 0}}

\newcommand{\Am}{{\bf A}}
\newcommand{\Bm}{{\bf B}}

\newcommand{\Em}{{\bf E}}
\newcommand{\Fm}{{\bf F}}

\newcommand{\Hm}{{\bf H}}

\newcommand{\Nm}{{\bf N}}

\newcommand{\Sm}{{\bf S}}

\newcommand{\Um}{{\bf U}}

\newcommand{\Vm}{{\bf V}}

\newcommand{\Ym}{{\bf Y}}
\newcommand{\Zm}{{\bf Z}}

\newcommand{\Cc}{{\cal C}}

\newcommand{\Ec}{{\cal E}}

\newcommand{\Gc}{{\cal G}}

\newcommand{\Ic}{{\cal I}}

\newcommand{\Lc}{{\cal L}}

\newcommand{\Nc}{{\cal N}}

\newcommand{\Sc}{{\cal S}}

\newcommand{\Uc}{{\cal U}}

\newcommand{\nuv}{\hbox{\boldmath$\nu$}}

\newcommand{\phiv}{\hbox{\boldmath$\phi$}}

\newcommand{\Sigmam}{\hbox{\boldmath$\Sigma$}}

\newcommand{\trace}{{\hbox{tr}}}

\newcommand{\eqdef}{\stackrel{\Delta}{=}}

\newcommand{\herm}{{\sf H}}

\newcommand{\SINR}{{\sf SINR}}
\newcommand{\SNR}{{\sf SNR}}

\usepackage{stmaryrd} 

\begin{document}

\setlength{\abovedisplayskip}{1pt}
\setlength{\belowdisplayskip}{1pt}
\setlength{\abovedisplayshortskip}{1pt}
\setlength{\belowdisplayshortskip}{1pt}
\allowdisplaybreaks

\title{Robust PCA for Subspace Estimation in User-Centric Cell-Free Wireless Networks}

\author{\IEEEauthorblockN{Fabian G\"ottsch\IEEEauthorrefmark{1},
		Noboru Osawa\IEEEauthorrefmark{2}, Takeo Ohseki\IEEEauthorrefmark{2}, Kosuke Yamazaki\IEEEauthorrefmark{2}, Giuseppe Caire\IEEEauthorrefmark{1}}
	\IEEEauthorblockA{\IEEEauthorrefmark{1}Technical University of Berlin, Germany\\
		\IEEEauthorrefmark{2}KDDI Research Inc., Japan\\
		Emails: \{fabian.goettsch, caire\}@tu-berlin.de, \{nb-oosawa, ohseki, ko-yamazaki\}@kddi-research.jp}}

\maketitle


\begin{abstract}
We consider a scalable user-centric cell-free massive MIMO network with distributed remote radio units (RUs), enabling macrodiversity and joint processing. 
Due to the limited uplink (UL) pilot dimension, multiuser interference in the UL pilot transmission phase makes channel estimation a non-trivial problem. 
We make use of two types of UL pilot signals, sounding reference signal (SRS) and demodulation reference signal (DMRS) pilots, for the estimation of the channel subspace and its instantaneous realization, respectively. 
The SRS pilots are transmitted over multiple time slots and resource blocks according to a Latin squares based hopping scheme, which aims at averaging out the interference of different SRS co-pilot users.
We propose a robust principle component analysis approach for channel subspace estimation from the SRS signal samples, employed at the RUs for each associated user. The estimated subspace is further used at the RUs for DMRS pilot decontamination and instantaneous channel estimation. We provide numerical simulations to
compare the system performance using our subspace and channel estimation scheme
with the cases of ideal partial subspace/channel knowledge and pilot matching channel estimation. The results show that a system with a properly designed SRS pilot hopping scheme can closely approximate the performance of a genie-aided system.
\end{abstract}

\vspace{-.2cm}
\section{Introduction} 

{\em Multiuser MIMO} and Marzetta's {\em massive MIMO} are key transformative technologies at the center of the last decade of theoretical research and practical system design \cite{caire2003achievable, 3gpp38211, marzetta2010noncooperative}.
Based on this, the joint processing of spatially distributed remote radio units (RUs) has been investigated more recently in numerous works (see \cite{9336188} for an overview).
We consider such a scalable user-centric {\em cell-free massive MIMO} network with $L$ remote RUs, each with $M$ antennas, and $K$ single-antenna user equipments (UEs), where finite size clusters of RUs are connected to each UE. 
We study the uplink (UL) pilot and data transmission, and use the received sounding reference signal (SRS) and demodulation reference signal (DMRS) pilots at the RUs for the estimation of the channel subspace and realization \cite{3gpp38211}. We note that, compared to the channel realization, the channel subspace changes on a much larger time scale and is constant over multiple resource blocks, and thus separate the assignment of SRS and DMRS pilots, maintaining the same RU clusters associated to the UEs. Although many works investigating cell-free wireless networks assume channel covariance matrices to be known throughout the system or at the RUs serving a UE (see \cite{9336188}), this knowledge is not easy to obtain in practical systems. 
On the other hand, in our previous work we showed that a simple projection of the DMRS pilot field onto the channel dominant subspace of the desired user is already sufficient to achieve significant pilot decontamination and closely approach the performance under ideal channel state information knowledge \cite{goettsch2021impact}. Therefore, we focus on the estimation of the channel dominant subspace, which is a more robust and less demanding task than the full knowledge of the channel covariance matrix.  
In this paper, we focus on an SRS pilot assignment scheme aiming to reduce pilot contamination and a subspace estimation approach, such that the estimated subspace can be used in place of the covariance matrix for further signal processing.

Subspace estimation based on Approximate Maximum-Likelihood is proposed in \cite{8264736} for a massive MIMO system, where the basic estimation problem for a single user without multiuser interference is studied.
Principle Component Analysis (PCA) for subspace estimation in a massive MIMO system is employed in \cite{9136836}, under the assumptions that no inter-cell interference occurs and that the base station (BS) receives signals from all users on the same multipaths. The estimated direction of arrival angles are thus the same for all users in the cell and only the channel gains are estimated per user with UL pilots.
In case of multiuser interference from different multipaths however, the received signals over time at the BS used for subspace estimation contain the UL signal from various users and possibly outliers, i.e., samples whose dominant subspace differs from the subspace of the desired signal. A PCA approach for datasets corrupted with outliers might lead to inaccurate estimates, which motivates the employment of {\em robust} PCA (R-PCA) that is able to identify these outliers and to find the underlying low-dimensional subspace of the inliers. We refer to \cite{8425657} for an overview of different R-PCA approaches.

In this work, all $K$ users send a sequence of $N$ SRS pilots in the UL. If $N<K$, SRS pilot collisions occur and the received SRS samples at the RUs possibly contain outliers. We develop an SRS pilot sequence assignment based on mutually orthogonal Latin squares \cite{467960} aiming to reduce outliers, and propose the R-PCA algorithm from \cite{6126034} at the RUs for subspace estimation of their associated users. While we have assumed perfect subspace knowledge in \cite{goettsch2021impact, kddi_uldl_precoding}, we show in this paper that the proposed schemes lead to accurate subspace estimates being able to closely approach the system performance assuming perfect subspace knowledge and ideal partial channel state information (CSI), respectively.

\section{System model}
We consider a cell-free wireless network in TDD operation with $L$ RUs, each with $M$ antennas, and $K$ single-antenna UEs. The set $\Cc_k \subseteq [L] = \{1,2,\dots, L\}$ denotes the cluster of RUs serving UE $k$ and the set $\Uc_\ell \subseteq [K]$ the cluster of UEs connected to RU $\ell$, where the cluster formation and DMRS pilot assignment for channel estimation follow the schemes in \cite{goettsch2021impact}. The RU-UE associations are described by a bipartite graph $\Gc$ whose vertices are the RUs and UEs, respectively. The set of edges accounting for associated RU-UE pairs is denoted by $\Ec$, i.e., $\Gc = \Gc([L], [K], \Ec)$. We consider OFDM modulation and channels following the standard block-fading model \cite{marzetta2010noncooperative,9336188,9064545}, such that they are random but constant over coherence blocks of $T$ signal dimensions in the time-frequency domain. 

We denote by $\HH \in \CC^{LM \times K}$ the channel matrix between all $LM$ RU antennas and all $K$ UE antennas on a given RB, where the $M \times 1$ block $\hv_{\ell,k}$ in the position corresponding to RU $\ell$ and UE $k$ denotes the channel between this RU-UE pair. We consider the {\em ideal partial CSI} regime, where each RU has perfect channel knowledge for its associated UEs. The known channel matrix of a cluster $\Cc_k$ is denoted by $\HH(\Cc_k) \in \CC^{LM \times K}$,
whose $M \times 1$ blocks of RU-UE pairs $(\ell,j) \in \Ec$ with $\ell \in \Cc_k$ are equal to $\hv_{\ell,j}$, and equal to $\zerov$ otherwise. 
The individual channels between RUs and UEs follow the simplified single ring local scattering model \cite{adhikary2013joint}, and $\Fm$ denotes the $M \times M$ unitary DFT matrix with $(m,n)$-elements
$\Fm_{m,n} = \frac{e^{-j\frac{2\pi}{M} mn}}{\sqrt{M}}$ for  $m, n  = 0,1,\ldots, M-1$.
Then, the channel between RU $\ell$ and UE $k$ is
\begin{equation} 
	\hv_{\ell,k} = \sqrt{\frac{\beta_{\ell,k} M}{|\Sc_{\ell,k}|}}  \Fm_{\ell,k} \nuv_{\ell, k},
\end{equation}
where $\Sc_{\ell,k} \subseteq \{0,\ldots, M-1\}$, $\nuv_{\ell,k}$ and $\beta_{\ell,k}$ are the angular support set according to \cite{adhikary2013joint}, an $|\Sc_{\ell,k}| \times 1$ i.i.d. Gaussian vector with components 
$\sim \Cc\Nc(0,1)$, and the large
scale fading coefficient (LSFC) including
pathloss, blocking effects and shadowing, respectively. 
Using a Matlab-like notation, $\Fm_{\ell,k} \eqdef \Fm(: , \Sc_{\ell,k})$ denotes the tall unitary matrix obtained by selecting the columns of $\Fm$ corresponding to the index set $\Sc_{\ell,k}$.

\subsection{Uplink data transmission} 
Let all UEs transmit with the same average energy per symbol $P^{\rm ue}$ (usually referred to as “power”), and we define the system parameter $\SNR \eqdef P^{\rm ue}/N_0$, 
where $N_0$ denotes the complex baseband noise power spectral density.
The received $LM \times 1$ symbol vector at the $LM$ RU antennas for a single channel use of the UL is given by
\begin{equation} 
	\yy^{\rm ul} = \sqrt{\SNR} \; \HH \sss^{\rm ul}   + \zz^{\rm ul}, \label{ULchannel}
\end{equation}
where $\sss^{\rm ul} \in \CC^{K \times 1}$ is the vector
of information symbols transmitted by the UEs (zero-mean unit variance and mutually independent random variables) and 
$\zz^{\rm ul}$ is an i.i.d. noise vector with components $\sim \Cc\Nc(0,1)$.  
The goal of cluster $\Cc_k$ is to produce an effective channel observation for symbol $s^{\rm ul}_k$ 
(the $k$-th component of the vector $\sss^{\rm ul}$ from the collectively received signal at the RUs $\ell \in \Cc_k$).  
We  define the combining coefficient $w_{\ell,k}$ of RU-UE pair $(\ell,k)$ and the receiver {\em unit norm} vector $\vvv_k \in \CC^{LM \times 1}$ formed by $M \times 1$ blocks
$w_{\ell,k} \vv_{\ell,k} : \ell = 1, \ldots, L$, such that $\vv_{\ell,k} = \zerov$ (the identically zero vector) if $(\ell,k) \notin \Ec$.
This reflects the fact that only the RUs in $\Cc_k$ are involved in producing a received observation for the detection of user $k$. 
The non-zero blocks $w_{\ell,k} \vv_{\ell,k} :  \ell \in \Cc_k$ contain the receiver combining vectors, where 
we use a local linear MMSE (LMMSE) principle for the computation of $\vv_{\ell,k}$,
and the coefficients $w_{\ell,k}$ are optimized by cluster $\Cc_k$ to maximize the UL SINR (see \cite{goettsch2021impact} for details).
The corresponding scalar combined observation for symbol $s^{\rm ul}_k$ is given by 
$
	r^{\rm ul}_k  = \vvv_k^\herm \yy^{\rm ul}. 
$
%
%
%
For simplicity, we assume that the channel decoder has perfect knowledge of the exact UL SINR value 
\begin{eqnarray} 
	\SINR^{\rm ul}_k 
	& = & \frac{  |\vvv_k^\herm \hh_k|^2 }{ \SNR^{-1}  + \sum_{j \neq k} |\vvv_k^\herm \hh_j |^2 },  \label{UL-SINR-unitnorm}
\end{eqnarray}
where $\hh_k$ denotes the $k$-th column of $\HH$. The corresponding UL {\em optimistic ergodic} achievable rate is given by 
\begin{equation}
	R_k^{\rm ul} = \EE [ \log ( 1 + \SINR^{\rm ul}_k ) ], \label{ergodic-rate}
\end{equation}
where the expectation is with respect to the small scale fading, while conditioning on the placement of UEs and RUs, and on the cluster formation.

\section{UL channel estimation}
\vspace{-.18cm}
In practice, ideal (although partial) CSI is not available and the channels $\{\hv_{\ell,k} : (\ell,k) \in \Ec\}$ must be estimated from
the orthogonal UL DMRS pilot sequences.  The DMRS pilot field received at RU $\ell$ is given by the $M \times \tau_p$ matrix of received symbols
\begin{equation} 
	\Ym_\ell^{\rm DMRS} = \sum_{i=1}^K \hv_{\ell,i} \phiv_{t_i}^\herm + \Zm_\ell^{\rm DMRS}, \label{Y_pilot} 
\end{equation}
where $\Zm_\ell^{\rm DMRS}$ is AWGN with elements i.i.d. $\sim \Cc\Nc(0, 1)$, and $\phiv_{t_i}$ denotes the DMRS pilot vector of dimension $\tau_p$ used by UE $i$ in the current RB, with  
total energy $\| \phiv_{t_i} \|^2 = \tau_p \SNR$.  For each UE $k \in \Uc_\ell$, RU $\ell$ produces the {\em pilot matching} (PM) channel estimates
\begin{align} 
	\widehat{\hv}^{\rm pm}_{\ell,k} =  \frac{1}{\tau_p \SNR} \Ym^{\rm DMRS}_\ell \phiv_{t_k}  
	= \hv_{\ell,k}  + \sum_{\substack{i : t_i = t_k \\ i\neq k}} \hv_{\ell,i}  + \widetilde{\zv}_{t_\ell,k} ,  \label{chest}
\end{align} 
where $\widetilde{\zv}_{t_\ell,k}$ has i.i.d. components $\Cc\Nc(0, \frac{1}{\tau_p\SNR})$. 
Notice that the presence of UEs $i \neq k$ using the same DMRS pilot $t_k$ yields pilot contamination, while $\phiv_{t_i}^\herm \phiv_{t_k} = 0, \ t_i \neq t_k$. 

Assuming that the subspace information $\Fm_{\ell,k}$ of all $k \in \Uc_\ell$ is known, 
we consider also the {\em subspace projection} (SP)  pilot decontamination scheme for which the projected channel estimate is given by the 
orthogonal projection of $\widehat{\hv}^{\rm pm}_{\ell,k}$ onto the subspace spanned by the columns of $\Fm_{\ell,k}$, i.e., 
\begin{align}
	\widehat{\hv}^{\rm sp}_{\ell,k} &= \Fm_{\ell,k}\Fm_{\ell,k}^\herm \widehat{\hv}^{\rm pm}_{\ell,k} 
	\label{chest1}
\end{align}
The pilot contamination term after the subspace projection  
is a Gaussian vector with mean zero and covariance matrix
\begin{equation}
	\Sigmam_{\ell,k}^{\rm co}  = \sum_{\substack{i : t_i = t_k \\ i\neq k}} \frac{\beta_{\ell,i} M}{|\Sc_{\ell,i}|}  \Fm_{\ell,k} \Fm_{\ell,k}^\herm \Fm_{\ell,i} \Fm^\herm_{\ell,i} \Fm_{\ell,k} \Fm^\herm_{\ell,k}. 
\end{equation}
When $\Fm_{\ell,k}$ and $\Fm_{\ell,i}$ are nearly mutually orthogonal, i.e. $\Fm_{\ell,k}^\herm \Fm_{\ell.i} \approx \zerov$,
the subspace projection is able to significantly reduce the DMRS pilot contamination effect.

\section{Subspace estimation with UL SRS pilots}

In order to implement the DMRS channel estimation with pilot decontamination based on channel subspace projection in (\ref{chest1}), 
the dominant subspace of each user channel $\hv_{\ell,k}$ with $k \in \Uc_\ell$ must be estimated at each RU $\ell$. Such subspaces are long-term statistical properties 
that depend on the geometry of the propagation, which is assumed to remain essentially invariant over sequences of many consecutive slots, 
and it is frequency-independent. Furthermore, assuming uniform linear arrays (ULAs) or uniform planar arrays (UPAs) and owing to the Toeplitz or block-Toeplitz structure of the channel covariance matrices, 
for large $M$ such subspaces are nearly spanned by subsets of the DFT columns (see \cite{adhikary2013joint}). In particular, this means that, while the one-ring scattering model may seem restrictive, it is actually general enough for large ULAs and UPAs, which is the most interesting 
and practical case for applications. 

In this section, we propose to use SRS pilots channel subspace estimation, where the sequence of SRS pilots spans a large number of slots and subcarriers in 
order to provide the necessary statistical averaging and capture the correct channel second-order statistics. We assume that each user sends just {\em one} SRS pilot symbol 
per slot, hopping over the subcarriers in multiple slots. In this sense, ``orthogonal SRS pilots'' indicate symbols sent by different users on different subcarriers on the same slot. 
We reserve a grid of $N$ distinct subcarriers for the SRS pilot in each slot, such that in any time slot there exist 
$N$ orthogonal SRS pilots.  We say that two users have an SRS pilot collision at some slot, when they transmit their SRS pilot symbol on the same subcarrier. 
The allocation of SRS pilot sequences to users is done in order to minimize the pilot collisions between users at the RUs forming their clusters. 
It is clear that two users physically separated by a large distance can have pilot collisions without suffering from the mutual interference, since their corresponding clusters are also
physically separated. In contrast, it may happen that some user $k$ collides on some slot $s$ with some user $j$, and that user $j$ is much closer to  RU $\ell \in \Cc_k$ than user $k$. In this case, the measurement at RU $\ell$ relative to user $k$ is heavily interfered by such collision. 
Our scheme is based on a) allocating sequences such that such type of damaging collision is minimized, and b) using a subspace estimation method that is robust to 
a small number of heavily interfered measurements (outliers). 
\subsection{Orthogonal Latin squares-based SRS pilot hopping scheme} 
We employ an SRS pilot hopping scheme based on orthogonal Latin squares \cite{467960}. 
A Latin square of order $N$ is an $N \times N$ array $\Am$ with elements in $[N] = \{1,\ldots, N\}$ such that each row and column contains all distinct elements. 
Two Latin squares $\Am$ and $\Bm$ are said to be mutually orthogonal if the collection of elementwise pairs 
$\{(\Am(i,j), \Bm(i,j)) : i,j \in [N]\}$ contains all $N^2$ distinct pairs. This means that the positions marked with $k \in [N]$ in $\Am$ are marked with all
possible distinct integers from 1 to $N$ in $\Bm$. 

\begin{example}  \label{latin-square}
	Consider $N = 5$ and the Latin squares
	\begin{equation}
		\Am = \Resize{.3\linewidth}{\begin{bmatrix}
				\boxed{1} & 2 & 3 & 4 & 5 \\
				2 & 3 & 4 & 5 & \boxed{1} \\
				3 & 4 & 5 & \boxed{1} & 2  \\
				4 & 5 & \boxed{1} & 2 & 3 \\
				5 & \boxed{1} & 2 & 3 & 4
			\end{bmatrix}}  , \
			\Bm = \Resize{.3\linewidth}{\begin{bmatrix}
				\boxed{1} & 2 & 3 & 4 & 5  \\
				3 & 4 & 5 & 1 & \boxed{2} \\
				5 & 1 & 2 & \boxed{3} & 4  \\
				2 & 3 & \boxed{4} & 5 & 1  \\
				4 & \boxed{5} & 1 & 2 & 3 
			\end{bmatrix} }. \nonumber
	\end{equation}
	The positions marked by (say) 1 in $\Am$ (boxed positions) contain all elements $\{1,2,3,4,5\}$ in $\Bm$, and this holds for all symbols. Hence, $\Am$ and $\Bm$ are mutually orthogonal Latin squares. \hfill $\lozenge$
\end{example}

The construction of families of $N - 1$ mutually orthogonal Latin squares of order $N$ when $N$ is a prime power (i.e., $N = p^n$ for some prime $p$ and integer $n$) 
is well-known and given in \cite{467960}. In the proposed scheme, a Latin square identifies $N$ hopping schemes as follows: we identify the rows of the 
Latin square with the distinct subcarriers $f \in [N]$ used for SRS hopping (numbered from 1 to N without loss of generality) and 
the columns with the  time slots $s \in [N]$. 
Each Latin square defines $N$ mutually orthogonal hopping sequences given by the sequence of positions $(f,s)$ corresponding to each integer $k \in [N]$. 
For example, in Example \ref{latin-square}, $N = 5$ users (say $k_1, k_2, \ldots, k_5$) are associated to the 5 mutually orthogonal hopping sequences
$\{ (f,s) : \Am(f,s) = n \}$, for $n = 1,2,\ldots, 5$ (e.g., user $k_1$ is associated to the sequence hopping over the boxed symbols in $\Am$ corresponding to index $n = 1$, and so on). 
If two users are associated to sequences in distinct orthogonal Latin squares, they shall collide only in one hopping position. In other words, a user associated to Latin square $\Am$ 
will collide on the $N$ hopping slots with each one of the $N$ users associated to the orthogonal Latin square $\Bm$. For example, consider the boxed positions in $\Am$ and $\Bm$
in Example \ref{latin-square}, where user $k_1$ associated to $n = 1$ in $\Am$ collides with different users associated with hopping sequences $1, 5, 4, 3, 2$ 
in $\Bm$ (in the order of consecutive slots $s = 1,2,3,4,5$ indicating the columns of the Latin square array). 
This creates a good averaging of interference. In particular, we partition the network coverage area into 
hexagonal cells, and allocate the $N-1$ Latin squares in a classical reuse of order $N-1$, such that adjacent cells have distinct orthogonal Latin squares. 
The hexagonal cells have nothing to do with the user-centric clusters and are defined on a purely geometric basis. 
We conclude this section by saying that  in general the length of SRS hopping sequences $S$ may be larger than $N$. In this case, the sequence is periodically repeated. 
\subsection{Estimation via R-PCA}
Consider a generic RU $\ell$. The RU is aware of the SRS hopping sequence of all its associated users in $\Uc_\ell$. 
Focusing on some UE $k \in \Uc_\ell$, we describe now how the RU can estimate the channel subspace (column span of $\Fm_{\ell,k}$ in our simplified scattering model).
The RU collects all SRS pilot measurements corresponding to the hopping sequence of UE $k$. On slot $s \in [S]$ these are given by 
\begin{align}
	\yv_{\ell,k}^{\rm SRS}(s) 
	& =  \hv_{\ell,k} (s) + \sum_{i \neq k: t_i^{\rm SRS}(s) = t_k^{\rm SRS}(s) } \hv_{\ell,i} (s) + \widetilde{\zv}_{\ell,k}(s) \nonumber \\
	& =  \hv_{\ell,k} (s) + \sum_{\substack{i \neq k: \\ i \in \Ic^s_k(s) }} \hv_{\ell,i} (s) +  \sum_{\substack{i \neq k: \\ i \in \Ic^w_k(s) }} \hv_{\ell,i} (s) + \widetilde{\zv}_{\ell,k} (s) \nonumber \\
	& =  \hv_{\ell,k} (s) + \ev_{\ell,k} (s) +  \nv_{\ell,k} (s) ,
\end{align}
where $\widetilde{\zv}_{\ell,k}(s)$ is $M \times 1$ Gaussian i.i.d. with components $\Cc\Nc(0, \frac{1}{\SNR})$, where the condition
$t_i^{\rm SRS}(s) = t_k^{\rm SRS}(s)$ indicates that the hopping sequence of user $i$ and of user $k$ collide at slot $s$, and where
the sets $\Ic^s_k(s)$ and $\Ic^w_k(s)$ contain the UEs colliding with UE $k$ with strong and weak LSFCs with respect to RU $\ell$, respectively. 
The term  $\ev_{\ell,k} (s) = \sum_{i \neq k: i \in \Ic^s_k(s) } \hv_{\ell,i} (s)$ accounts for the strong undesired signals (the so-called \textit{outliers}) while the term
$\nv_{\ell,k} = \sum_{i \neq k: i \in \Ic^w_k(s) } \hv_{\ell,i} (s) + \widetilde{\zv}_{\ell,k} (s)$ includes noise and weak interference.
Stacking the $S$ SRS pilot observations as columns of an $M \times S$ array, we have
\begin{eqnarray}
	\Ym^{\rm SRS}_{\ell,k} & = & [\yv_{\ell,k}^{\rm SRS}(1) \ \yv_{\ell,k}^{\rm SRS}(2) \dots \yv_{\ell,k}^{\rm SRS}(S)] \label{Y_concat} \\
	& = & \Hm_{\ell,k} + \Nm_{\ell,k} + \Em_{\ell,k}, 
\end{eqnarray}
where $\Hm_{\ell,k}$, $\Nm_{\ell,k}$ and $\Em_{\ell,k}$ are given by the analogous stacking of vectors $\hv_{\ell,k} (s), \ev_{\ell,k} (s)$ and $ \nv_{\ell,k} (s)$, respectively. 
Because of the orthogonal Latin squares hopping patterns, it is expected that 
$\Em_{\ell,k}$ is column-sparse to a certain degree.  This is equivalent to the noise plus outliers model in \cite{6126034}. 
The R-PCA algorithm in  \cite{6126034} aims at detecting outliers, i.e., the non-zero columns of $\Em_{\ell,k}$, and at estimating the subspace of $\Hm_{\ell,k}$, which eventually is the desired channel subspace
of UE $k$ at RU $\ell$. Fixing some $\epsilon > 0$ and $\lambda > 0$, 
the algorithm solves the  convex problem
\begin{eqnarray}
	\underset{ \Hm_{\ell,k}, \Em_{\ell,k} }{\text{minimize}} & & \lVert \Hm_{\ell,k} \rVert_* + \lambda \lVert \Em_{\ell,k} \rVert_{2,1} \label{pca_opt_problem} \\
	\text{subject to:} & & \lVert \Ym^{\rm SRS}_{\ell,k} - \Hm_{\ell,k} - \Em_{\ell,k} \rVert_F \leq \epsilon, \nonumber
\end{eqnarray}
where $\lVert \cdot \rVert_*$, $\lVert \cdot \rVert_F$, and $\lVert \cdot \rVert_{2,1}$ are the nuclear norm, the Frobenius norm, and the sum of the $\ell_2$ column norms of a matrix, respectively. The Lagrangian function of (\ref{pca_opt_problem}) is given by 
\begin{align}
	\Lc &\left (\Hm_{\ell,k}, \Em_{\ell,k}, \lambda, \mu \right )   = \lVert \Hm_{\ell,k} \rVert_* \nonumber \\  &+ \lambda \lVert \Em_{\ell,k} \rVert_{2,1} + \mu \left ( \lVert \Ym^{\rm SRS}_{\ell,k} - \Hm_{\ell,k} - \Em_{\ell,k} \rVert_F - \epsilon \right ),
\end{align}
where $\mu$ is a Lagrange multiplier.
Therefore, the corresponding unconstrained convex minimization problem is given as
\begin{align}
	\underset{ \Hm_{\ell,k}, \Em_{\ell,k} }{\min}  \; \underset{\mu \geq 0}{\max} \hspace{.2cm} \Lc \left (\Hm_{\ell,k}, \Em_{\ell,k}, \lambda, \mu \right ). \label{pca_convex}
\end{align}
We employ the algorithm proposed in \cite{6126034} to approach (\ref{pca_convex}), which returns estimates $\widehat{\Hm}_{\ell,k}$ and $\widehat{\Em}_{\ell,k}$ 
of the channel and outliers matrix, respectively. 
The parameter $\lambda$ is given as an input to the algorithm and is optimized empirically. If the rank of the resulting $\widehat{\Hm}_{\ell,k}$ is larger (resp., smaller) than expected, we decrease (resp., increase) $\lambda$. The algorithm does not require $\epsilon$ to be specified and treats the observation matrix as in the noiseless case. Note that $\epsilon$ however is important for the analytical results of the algorithm in \cite{6126034}. The Lagrange multiplier $\mu$ is optimized by the algorithm via primal-dual iterations.

From the SVD $\widehat{\Hm}_{\ell,k} = \widehat{\Um}\widehat{\Sm} \widehat{\Vm}^\text{H}$, we estimate the subspace by considering the left singular vectors (columns of $\widehat{\Um}$) corresponding to the dominant singular values. One approach to find the number of dominant singular values is to find the index at which there is the largest difference (gap) between consecutive singular values (diagonal entries of $\widehat{\Sm}$).
Let $\widehat{\Fm}_{\ell,k}^{\rm PCA} = \widehat{\Um}_{:,1:r^{\rm PCA}}$ denote the tall unitary matrix obtained by selecting the dominant $r^{\rm PCA}$ left eigenvectors as explained above.  We can further post-process the subspace estimate by imposing that its basis vectors are DFT columns. 
As anticipated before, this is motivated by the fact that, for large Toeplitz matrices, the eigenvectors are closely approximated by DFT vectors \cite{adhikary2013joint}. 
The identification of the best fitting DFT columns to the R-PCA estimated subspace can be done greedily by finding one by one the $r^{\rm PCA}$ columns $\Fm_{:,i}$ of $\Fm$ 
that maximize the quantity
\begin{eqnarray}
	\Fm_{:,i}^\herm \widehat{\Fm}_{\ell,k}^{\rm PCA} (\widehat{\Fm}_{\ell,k}^{\rm PCA})^\herm \Fm_{:,i}.
\end{eqnarray}
We denote the selected set of column indices as $\widehat{\Sc}_{\ell,k}^{\rm PP}$, such that the corresponding estimated projected PCA (PP) 
subspace is given by $\widehat{\Fm}_{\ell,k}^{\rm PP} = \Fm(: , \widehat{\Sc}_{\ell,k}^{\rm PP})$. The estimated channel covariance matrix for a given subspace estimate 
$\widehat{\Fm}_{\ell,k}$ is given by 
\begin{eqnarray}
	\Sigma_{\hv}(\widehat{\Fm}_{\ell,k}) = \frac{\beta_{\ell,k} M}{r^{\rm PCA}} \widehat{\Fm}_{\ell,k} \widehat{\Fm}_{\ell,k}^\herm, 
\end{eqnarray} 
where $\widehat{\Fm}_{\ell,k} = \widehat{\Fm}_{\ell,k}^{\rm PCA}$ or $\widehat{\Fm}_{\ell,k} = \widehat{\Fm}_{\ell,k}^{\rm PP}$.

In order to evaluate the quality of a generic subspace estimate $\widehat{\Fm}_{\ell,k}$, we consider the power efficiency defined by 
\begin{eqnarray}
	E_{\rm PE}(\widehat{\Fm}_{\ell,k}) = \frac{\trace\left(\Sigma_{\hv}(\Fm_{\ell,k}) \Sigma_{\hv}(\widehat{\Fm}_{\ell,k}) \right)}{\trace\left(\Sigma_{\hv}(\Fm_{\ell,k}) \Sigma_{\hv}(\Fm_{\ell,k}) \right)},
\end{eqnarray}
where $\Sigma_{\hv}(\Fm_{\ell,k}) = \frac{\beta_{\ell,k} M}{|\Sc_{\ell,k}|} \Fm_{\ell,k} (\Fm_{\ell,k})^\herm$. 
The power efficiency $E_{\rm PE}(\widehat{\Fm}_{\ell,k}) \in [0,1]$ measures how much power from the desired channel in (\ref{chest1}) is captured in the channel estimate. 
\vspace{-.0cm}
\section{Simulations} \label{sec:ul_perfect_csi}
\vspace{-.0cm}
\begin{figure}[t!]
	\centerline{\includegraphics[width=4.4cm]{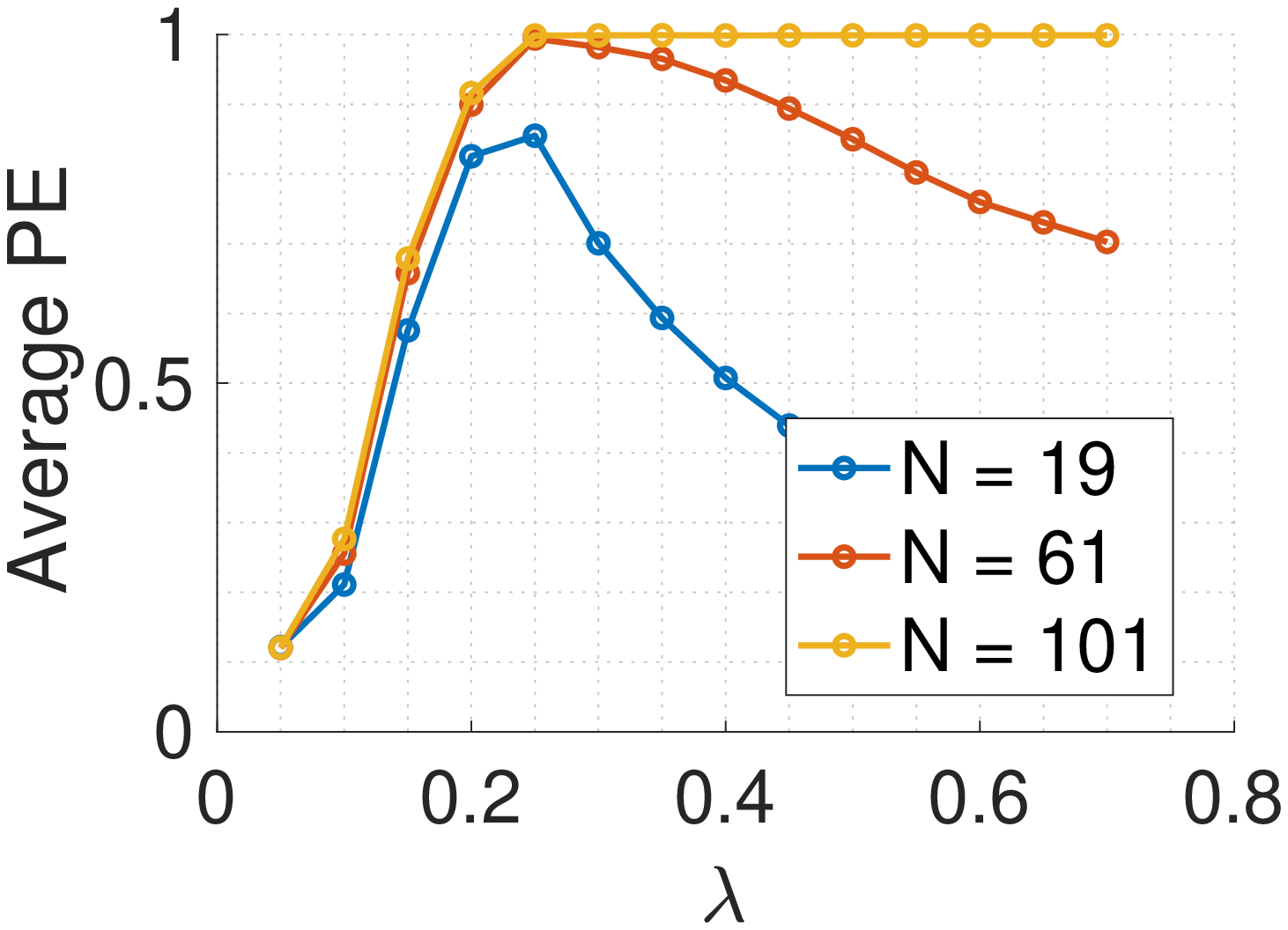}  \hspace{-.0cm}
		\includegraphics[width=4.4cm]{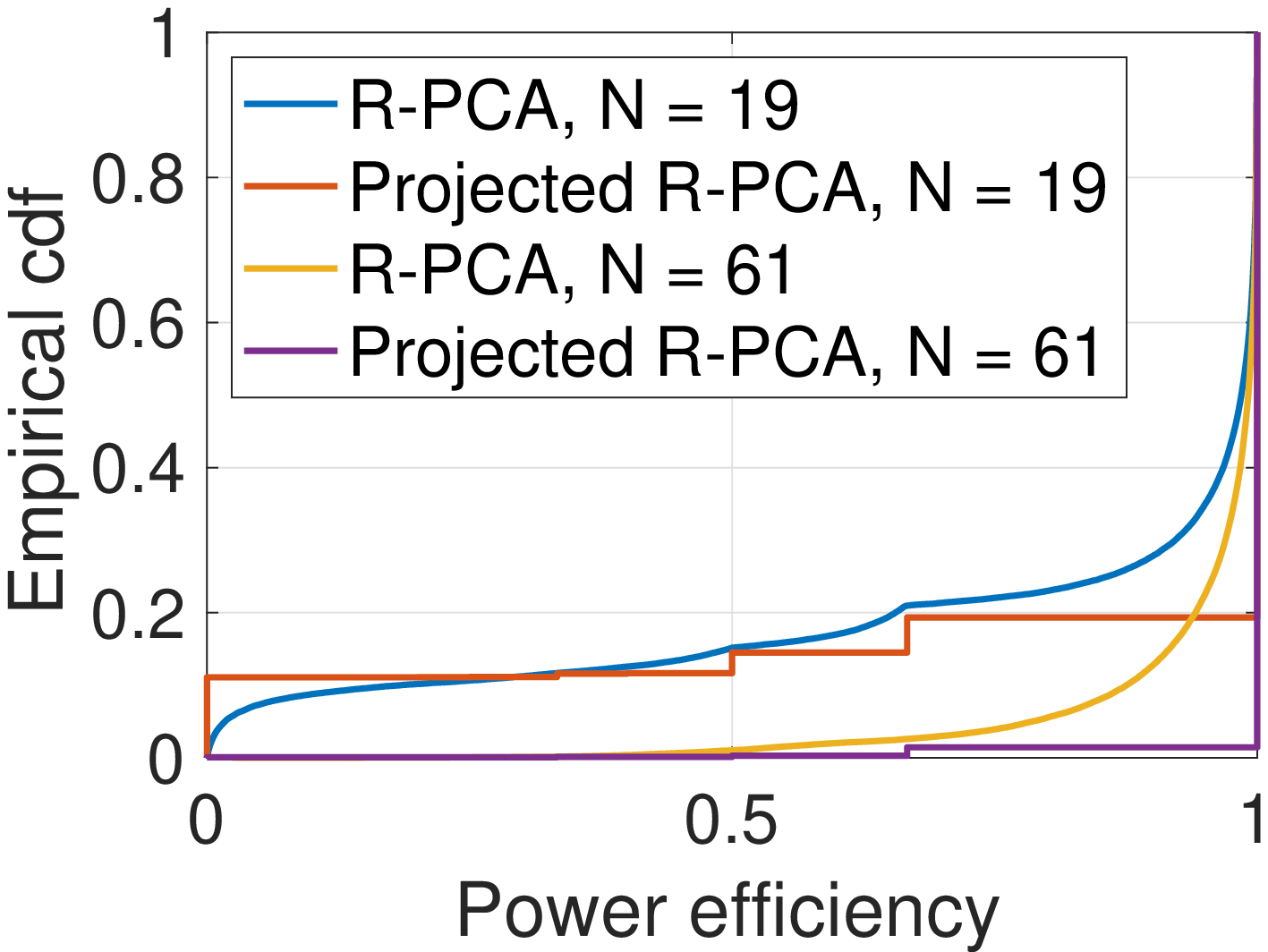}}
	\vspace{-.3cm}
	\caption{The average PE of the projected R-PCA subspace estimates with $L=40, M=16$ for different $N$ and $\lambda$ (left). The PE of the R-PCA subspace estimates with $L=40, M=16$, $\lambda=0.25$ and $N = \{19,61\}$ (right).}
	\label{subspace_estimation_results_pca}
\end{figure}

In our simulations, we consider a square coverage area of $A = 2 \times 2$ $\text{km}^2$ with a torus topology to avoid boundary effects. 
The LSFCs are given  according to the 3GPP urban microcell pathloss model from \cite{3gpp38901}. 
The average energy per symbol $P^{\rm ue}$ is chosen such that $\bar{\beta} M \SNR = 1$ (i.e., 0 dB), when the expected pathloss $\bar{\beta}$ with respect to LOS and NLOS is calculated for distance $3 d_L$, where $d_L = \sqrt{\frac{A}{\pi L}}$ is the radius of a disk of area equal to $A/L$.
We consider RBs of dimension $T = 200$ symbols, and the UL spectral efficiency (SE) for UE $k$ is given by  
\begin{equation}
	{\rm SE}^{\rm ul}_k =  (1 - \tau_p/T) R_k^{\rm ul}.
\end{equation}
The angular support $\Sc_{\ell,k}$ contains the DFT quantized angles (multiples of $2\pi/M$) falling inside an interval of length $\Delta$ placed symmetrically around the direction joining UE $k$ and RU $\ell$. We use $\Delta = \pi/8$ and the maximum cluster size $Q=10$ (RUs serving one UE) in the simulations. The SNR threshold $\eta=1$ makes sure that an RU-UE association can only be established, when $\beta_{\ell,k} \geq \frac{\eta}{M \SNR } $. For each studied system setup, we generate 100 different realizations (random uniform placement of RUs and UEs). Unless otherwise stated, we consider a system with $L=40$ RUs, $M=16$ RU antennas, $K=100$ UEs and $\tau_p = 15$.

\subsection{Subspace estimation accuracy}
The left plot of Fig. \ref{subspace_estimation_results_pca} shows the average PE of the projected subspace estimates for different $N$ and $\lambda$.
With $K<N=101$, the SRS pilots of different UEs never collide and the estimation accuracy is higher than for smaller $N$, where $K>N$. For $N=61$ and $\lambda = 0.25$ however, we can closely approach the accuracy of $N = 101$. The right plot of Fig. \ref{subspace_estimation_results_pca} shows the empirical cdf of the subspace estimates before and after projection onto the columns of the DFT matrix for $\lambda=0.25$ and $N = \{19,61\}$.
The results of the R-PCA outputs follow a continuous distribution, since the R-PCA estimates the subspace without the knowledge that it is constructed by a set of columns of the DFT matrix. 
The results of the post-processed estimates follow approximately the distribution of the R-PCA estimates, with the difference that due to the post-processing the set of possible outcomes is discrete. Thanks to less SRS pilot collisions, the fraction of estimates with PE equal to $1$ is larger with $N=61$ compared to $N=19$.


\subsection{System performance with subspace estimates}
\begin{figure}[t!]
	\centerline{\includegraphics[width=.49\linewidth]{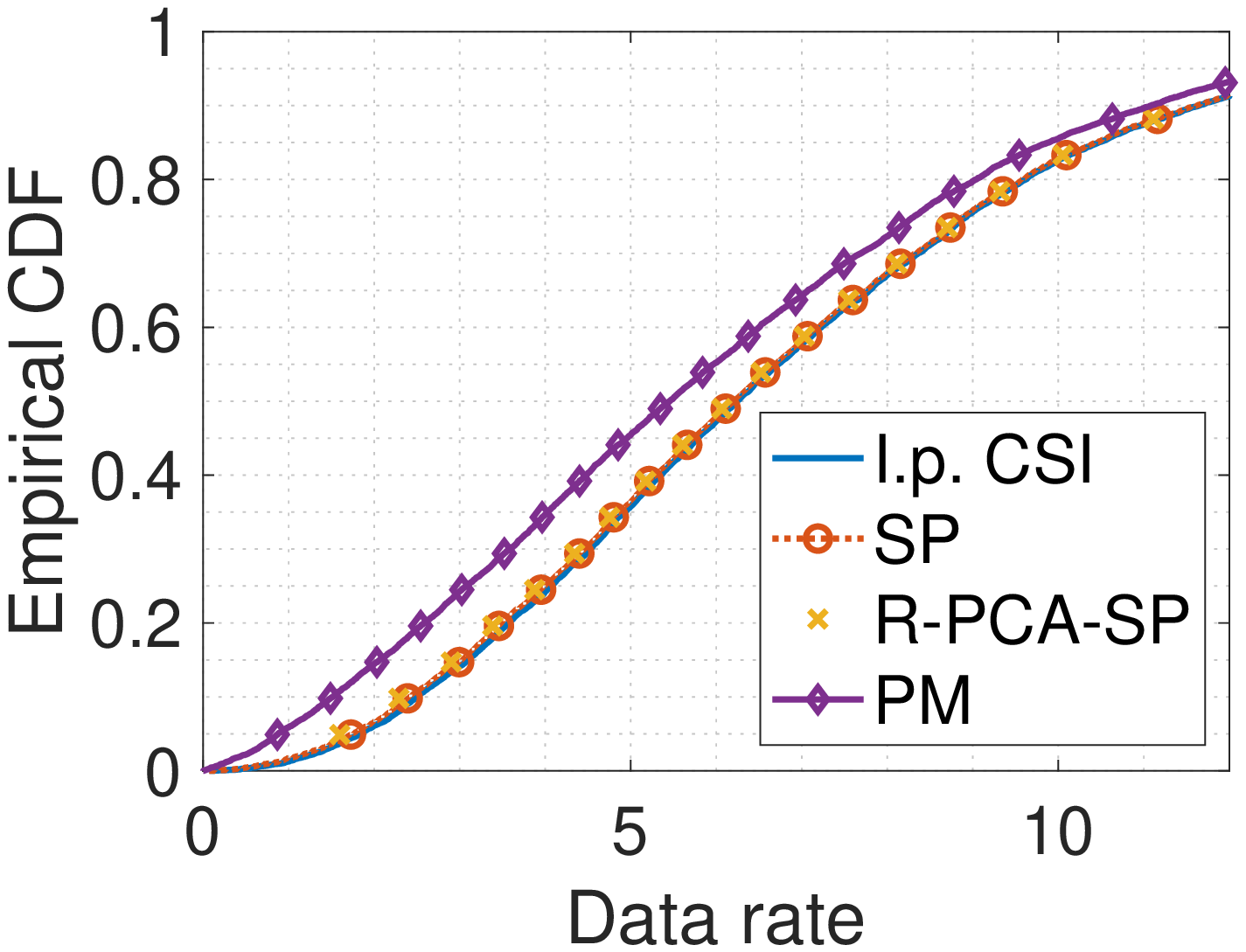}  
		\hspace{-.0cm}
		\includegraphics[width=.49\linewidth]{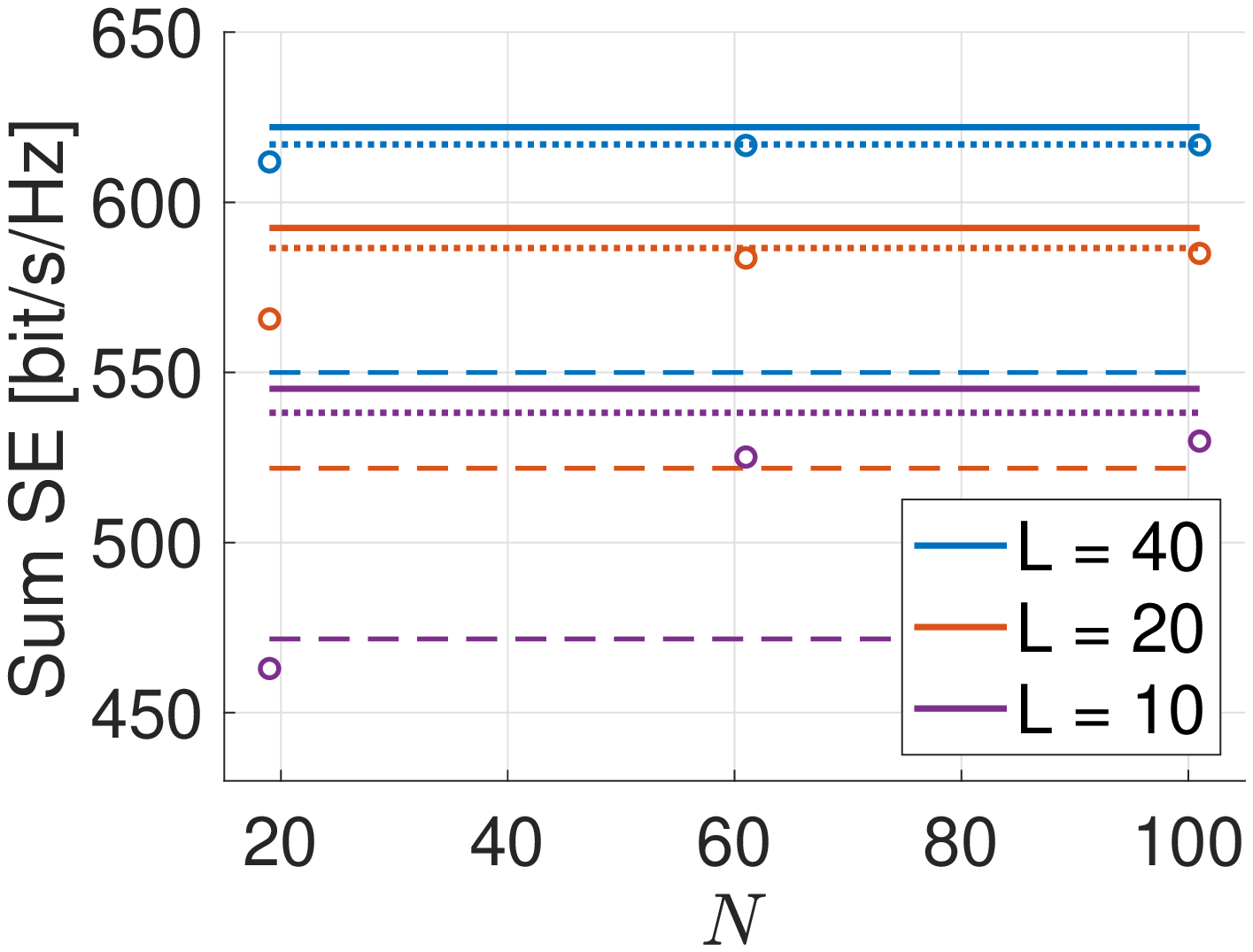}
	}
	\vspace{-.3cm}
	\caption{Left: The per-user UL data rates, where $N = 19$. Right: The sum SE achieved with i.p. CSI (solid lines), SP (dotted lines),  R-PCA-SP (circles) and PM (dashed lines).}
	\label{rate_evaluation_pca}
\end{figure}

We use the projected R-PCA estimates with $\lambda=0.25$ and $N = \{19,61,101\}$ for subspace projection channel estimation in (\ref{chest1}). 
We evaluate the achieved UL optimistic ergodic achievable rates in a cell-free network with the subspace estimates, and compare the results to the cases of ideal partial CSI, channel estimation with perfect subspace knowledge in (\ref{chest1}), and PM channel estimation, respectively.
For each set of parameters we generated 100 independent layouts, 
and for each  layout we computed the expectation in (\ref{ergodic-rate}) 
by Monte Carlo averaging with respect to the channel vectors. 
The left plot of Fig. \ref{rate_evaluation_pca} shows that the gap between the rates with perfect subspace knowledge (denoted by ``SP") and subspace estimates (``R-PCA-SP") can be basically closed. The rates with ideal partial CSI (``i.p. CSI") can be closely approached by the subspace projection channel estimation, while the system performance using PM is limited due to pilot contamination. 
Although Fig. \ref{subspace_estimation_results_pca} shows that a significant fraction of the subspace estimates has a PE less than 1, the almost inexisting gap between ``SP'' and ``R-PCA-SP" can be explained by the fact that bad estimates occur most likely for RU-UE pairs with a relatively small channel gain. The corresponding RUs contribute only to a little extent to the UE's data rate, such that a bad estimate does not degrade significantly the UE's data rate.

The right plot of Fig. \ref{rate_evaluation_pca} shows the sum SE for different levels of antenna distribution, where $LM=640$ for all choices of $L$. For each $L$, the parameters $\tau_p$ and $\lambda$ are chosen to maximize the sum SE. We observe an increasing gap between the sum SE with ideal partial CSI and the sum SE with R-PCA-SP for smaller $L$. We can explain this by the fact that for smaller $L$ each RU has to serve more UEs compared to larger $L$. On average, the channel gain of these RU-UE pairs will be smaller, leading to less accurate subspace estimates, which in turn degrade the instantaneous channel estimation. The overall better performance (independent of the CSI scenario) of distributed antenna configurations such as $L=40$ results from more RU-UE associations with large channel gains and increased macrodiversity.

\section{Conclusions}
The results show that the proposed SRS pilot assignment and subspace estimation schemes under practical assumptions can basically obtain the same performance as a system with perfect subspace information. Furthermore, channel estimation using subspace estimates yields data rates which approach very closely the performance of ideal partial CSI, pointing out the powerfulness of subspace information, especially under the aspect that subspace knowledge is less challenging to obtain than the covariance matrix. If appropriately designed, we can conclude that the proposed SRS hopping and R-PCA based subspace estimation essentially solve the problem of pilot contamination in cell-free TDD wireless networks.

\bibliography{IEEEabrv,spawc22-paper}

\end{document}